\documentclass{cernrep} 
\usepackage{texnames}
\usepackage[T1]{fontenc}
\begin{document}
\title{Statistical Challenges of Global SUSY Fits}
 
\author{Roberto Trotta$^1$ \& Kyle Cranmer$^2$}
\institute{$^1$Imperial College London. $^2$New York University}

\maketitle 

\begin{abstract}
We present recent results aiming at assessing the coverage properties of Bayesian and frequentist inference methods, as applied to the reconstruction of supersymmetric parameters from simulated LHC data. We discuss the statistical challenges of the reconstruction procedure, and highlight the algorithmic difficulties of obtaining accurate profile likelihood estimates.
\end{abstract}
 
\section{Introduction}
 
Experiments at the Large Hadron Collider (LHC) have already started testing
many models of particle physics beyond the Standard Model (SM), and particular attention is being paid to the Minimal Supersymmetric SM
(MSSM) and to other scenarios involving softly-broken
supersymmetry (SUSY).

In the last few years, parameter inference methodologies have been developed, applying both Frequentist and Bayesian
statistics (see e.g.,~\cite{Baltz:2004aw,Allanach:2005kz,deAustri:2006pe,Allanach:2007qk,Roszkowski:2007fd, Buchmueller:2009fn}). While the
efficiency of Markov Chain Monte Carlo (MCMC) techniques has allowed for a full exploration of
multidimensional models, the likelihood function from present data  is multimodal with many narrow features, making the
exploration task with conventional MCMC methods challenging. A powerful alternative to classical MCMC has emerged in the form of Nested Sampling~\cite{skilling04}, a Monte Carlo method whose primary aim is the efficient calculation of the Bayesian evidence, or model likelihood. As a by-product, the algorithm also produces samples from the posterior distribution. Those same samples can also be used to estimate the profile likelihood. {\sc MultiNest} \cite{multinest}, a publicly available implementation of the nested sampling algorithm, has been shown to reduce the computational
cost of performing Bayesian analysis typically by two orders of magnitude as compared with basic MCMC techniques. {\sc MultiNest} has been integrated in the
\texttt{SuperBayeS} code\footnote{Available from: \texttt{www.superbayes.org}} for fast and efficient exploration of SUSY models.

Having implemented sophisticated statistical and scanning methods, 
several groups have turned their attention to
evaluating the sensitivity to the choice of priors
\cite{Allanach:2007qk,Lafaye:2007vs,Trotta:2008bp} and of scanning
algorithms \cite{Akrami:2009hp}.  Those analyses indicate that current constraints are not strong enough to dominate the Bayesian posterior and that
the choice of prior does influence the resulting inference.  While
confidence intervals derived from the profile likelihood or a
chi-square have no formal dependence on a prior, there is a sampling
artifact when the contours are extracted from samples produced from
Bayesian sampling schemes, such as MCMC or {\sc MultiNest}~\cite{Trotta:2008bp}.

Given the sensitivity to priors and the differences between the
intervals obtained from different methods, it is natural to seek out a
quantitative assessment of their performance, namely their \textit{coverage}: the probability that an interval
will contain (cover) the true value of a parameter.  The defining
property of a 95\% confidence interval is that the procedure adopted for its estimation should produce intervals that cover the true value 95\% 
of the time; thus, it is reasonable to check if the
procedures have the properties they claim. 
While Bayesian techniques are not designed with coverage as a goal, it
is still meaningful to investigate their coverage properties. Moreover, the frequentist intervals
obtained from the profile likelihood or chi-square functions are based
on asymptotic approximations and are not guaranteed to have the
claimed coverage properties.

Here we report on recent studies investigating the coverage properties of both Bayesian and
Frequentist procedures commonly used in the literature. We also highlight the numerical and sampling challenges that have to be met in order to obtain a sufficienlty high-resolution mapping of the profile likelihood when adopting Bayesian algorithms (which are typically designed to map out the posterior mass, instead). 

For the sake of example, we consider in the following the so-called mSUGRA or
Constrained Minimal Supersymmetric Standard Model (CMSSM), a model with fairly strong universality 
assumptions regarding the SUSY breaking parameters, which reduce the number of free parameters to be estimated to just five, denoted by the symbol ${\bf \Theta}$: common scalar ($m_0$), gaugino ($m_{1/2}$) and
tri--linear ($A_0$) mass parameters (all specified at the GUT
scale) plus the ratio of Higgs vacuum expectation values $\tan\beta$ and
$\text{sign}(\mu)$, where $\mu$ is the Higgs/higgsino mass parameter
whose square is computed from the conditions of radiative electroweak
symmetry breaking (EWSB).

\section{Coverage study of the CMSSM}

\subsection{Accelerated inference from neural networks}

Coverage studies require extensive computational
expenditure, which would be unfeasible with standard
analysis techniques.  Therefore, in Ref.~\cite{Bridges:2010de} a class 
of machine learning devices called Artificial Neural Networks (ANNs) was used to approximate the most computationally intensive sections of the
analysis pipeline.

\begin{figure}
\begin{center}
\includegraphics[width=0.48\linewidth]{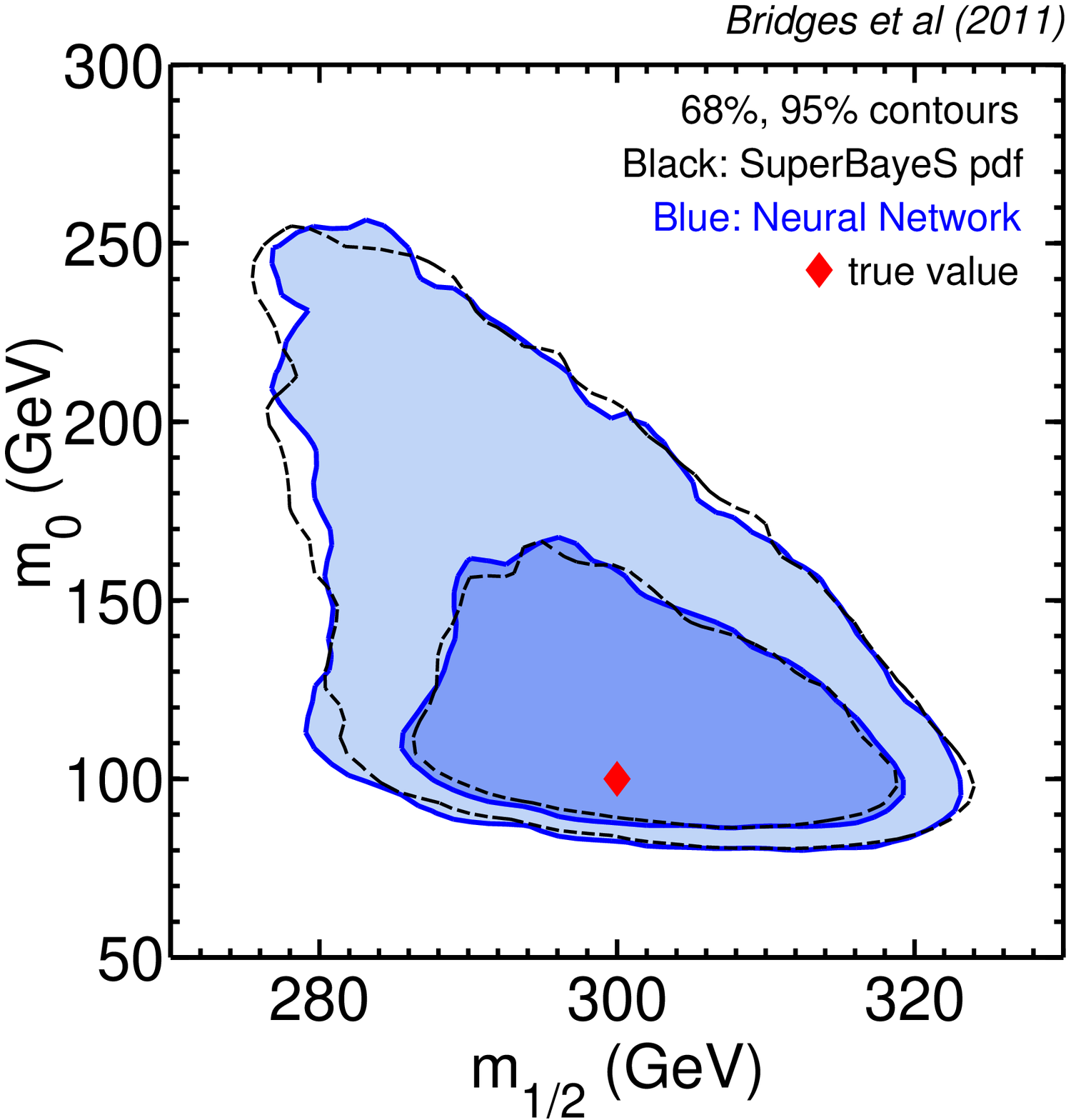}
\includegraphics[width=0.48\linewidth]{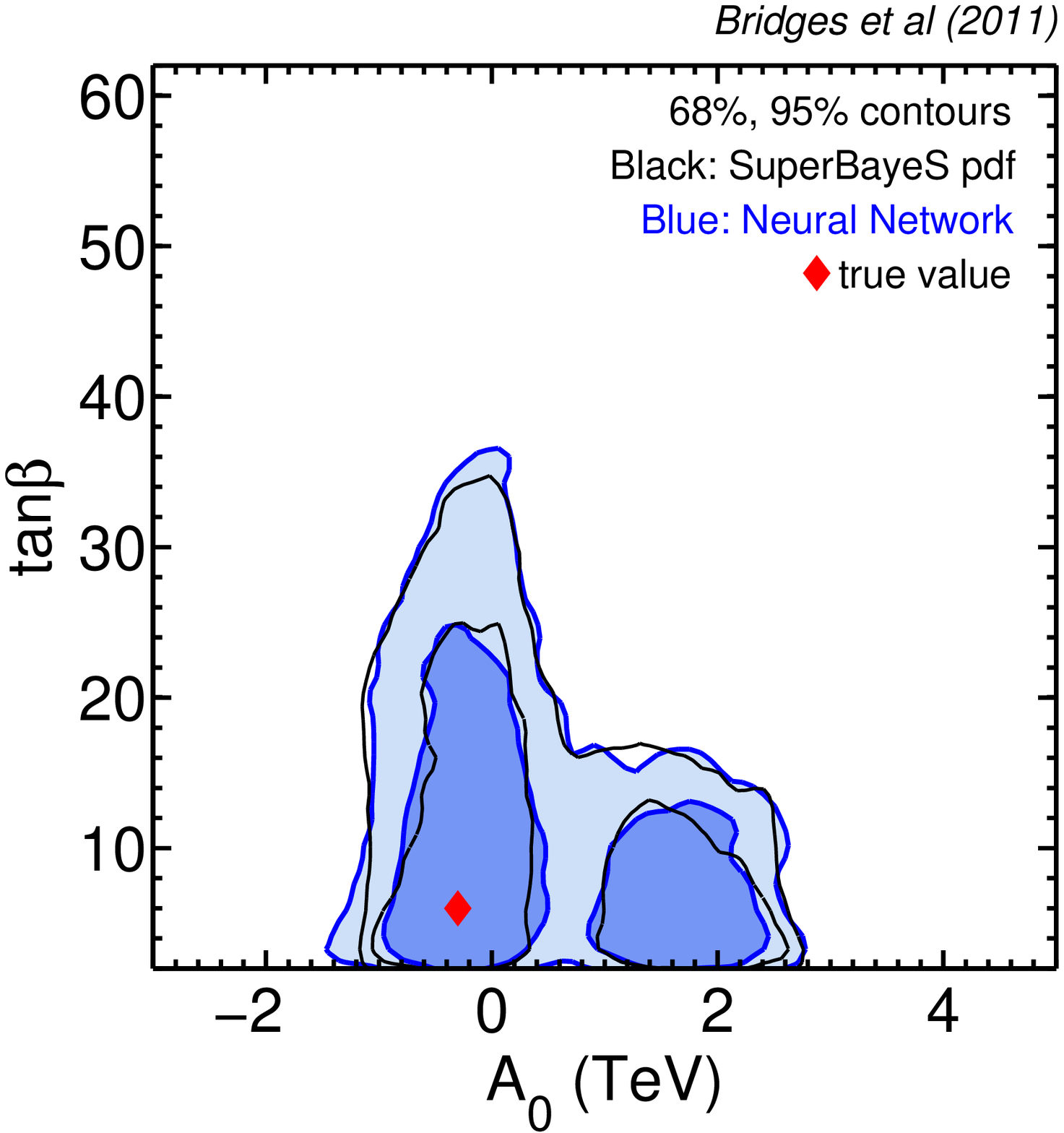}
\end{center}
\caption{\label{fig:nn_comparison} Comparison of Bayesian posteriors obtained by solving the RGEs fully numerically (black lines, giving 68\% and 95\% regions) and neural networks (blue lines and corresponding filled regions), from simulated ATLAS data. The red diamond gives the true value for the benchmark point adopted.  From~\cite{Bridges:2010de}.}
\end{figure}

Inference on the parameters of interest ${\bf \Theta}$ requires relating them to observable quantities, such as the sparticle
mass spectrum at the LHC, denoted by $\bf{m}$, over which the likelihood is defined. This is achieved by
evolving numerically the Renormalization Group Equations (RGEs) using publicly available codes, which is however a computationally intensive procedure.  One can view the RGEs simply as a mapping from ${\bf \Theta} \to
\bf{m}$, and attempt to engineer a computationally efficient
representation of the function. In~\cite{Bridges:2010de}, an adequate solution was provided by a three-layer perceptron, a type of feed-forward
neural network consisting of an input layer (identified with ${\bf \Theta}$), a hidden layer and an
output layer (identified with the value of ${\bf m}({\bf \Theta})$ that we are trying to approximate). The weight and biases defining the network were determined via an appropriate training procedure, involving the minimization of a loss function (here, the discrepancy between the value of  ${\bf m}({\bf \Theta})$ predicted by the network and its correct value obtained by solving the RGEs) defined over a set of 4000 training samples. A number of tests on the accuracy and noise of the network were performed, showing a correlation in excess of 0.9999 between the approximated value of ${\bf m}({\bf \Theta})$ and the value obtained by solving the RGEs for an independent sample. A second classification network was employed to distinguish between physical and un-physical points in parameter space (i.e., values of ${\bf \Theta}$ that do not lead to physically viable solutions to the RGEs). The final result of replacing the computationally expensive RGEs with the ANNs is presented in Fig.~\ref{fig:nn_comparison}, which shows that the agreement between the two methods is excellent, within numerical noise. By using the neural network, a speed-up factor of about $3 \times 10^4$ compared with scans using the explicit spectrum calculator was observed.

\subsection{Coverage results for the ATLAS benchmark}
\label{sec:coverage}

We studied the coverage properties of intervals obtained for 
the so-called ``SU3'' benchmark point. To this end, we need the ability to generate pseudo-experiments with ${\bf \Theta}$ fixed at the value of the benchmark. We adopted a parabolic approximation of the log-likelihood function (as reported in Ref.~\cite{atlas09}), based on the measurement of edges and thresholds in the invariant mass distributions for various combinations of leptons and jets in final state of the selected candidate SUSY events,  assuming an integrated 
luminosity of 1 $\text{fb}^{-1}$ for ATLAS. Note that the relationship between the sparticle masses and the directly observable mass edges is highly non-linear, so a Gaussian is likely to be a poor approximation to the actual likelihood function.  Furthermore, these edges share several sources of systematic uncertainties, such as jet and lepton energy scale uncertainties, which are only approximately communicated in Ref.~\cite{atlas09}.  Finally, we introduce the additional simplification that the likelihood is also a multivariate Gaussian with the same covariance structure.  We constructed $10^4$ pseudo-experiments and analyzed them with both MCMC (using a Metropolis-Hastings algorithm) and {\sc MultiNest}. Altogether, our neural network MCMC runs have performed a total of $4 \times 10^{10}$ likelihood evaluations, in a total computational effort of approximately  $2\times 10^4$ CPU-minutes. We estimate that the solving the RGEs fully numerically would have taken about 1100-CPU years, which is at the boundary of what is feasible today, even with a massive parallel computing effort. 

\begin{figure}
\begin{center}
\includegraphics[width=0.32\linewidth]{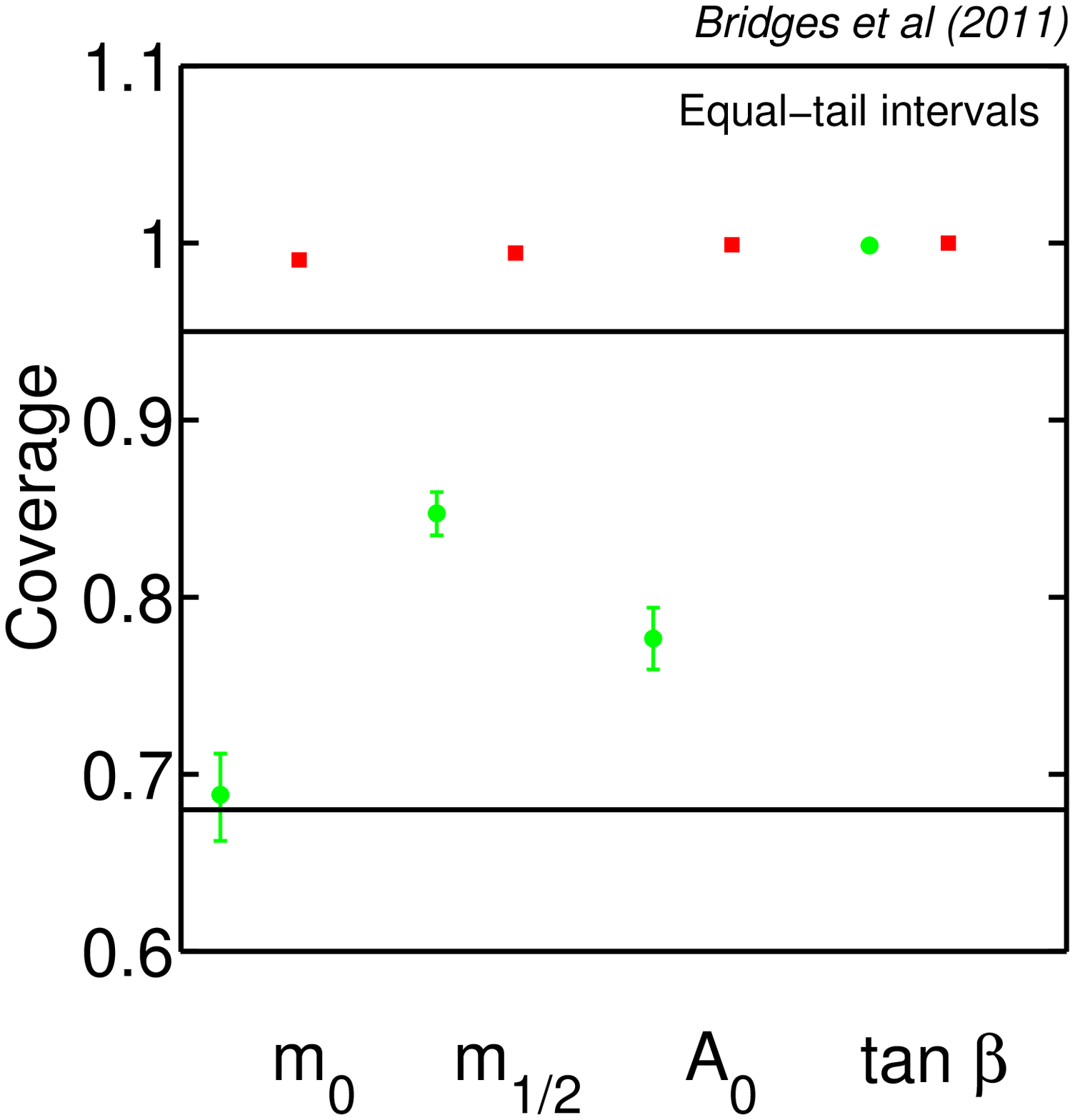}
\includegraphics[width=0.32\linewidth]{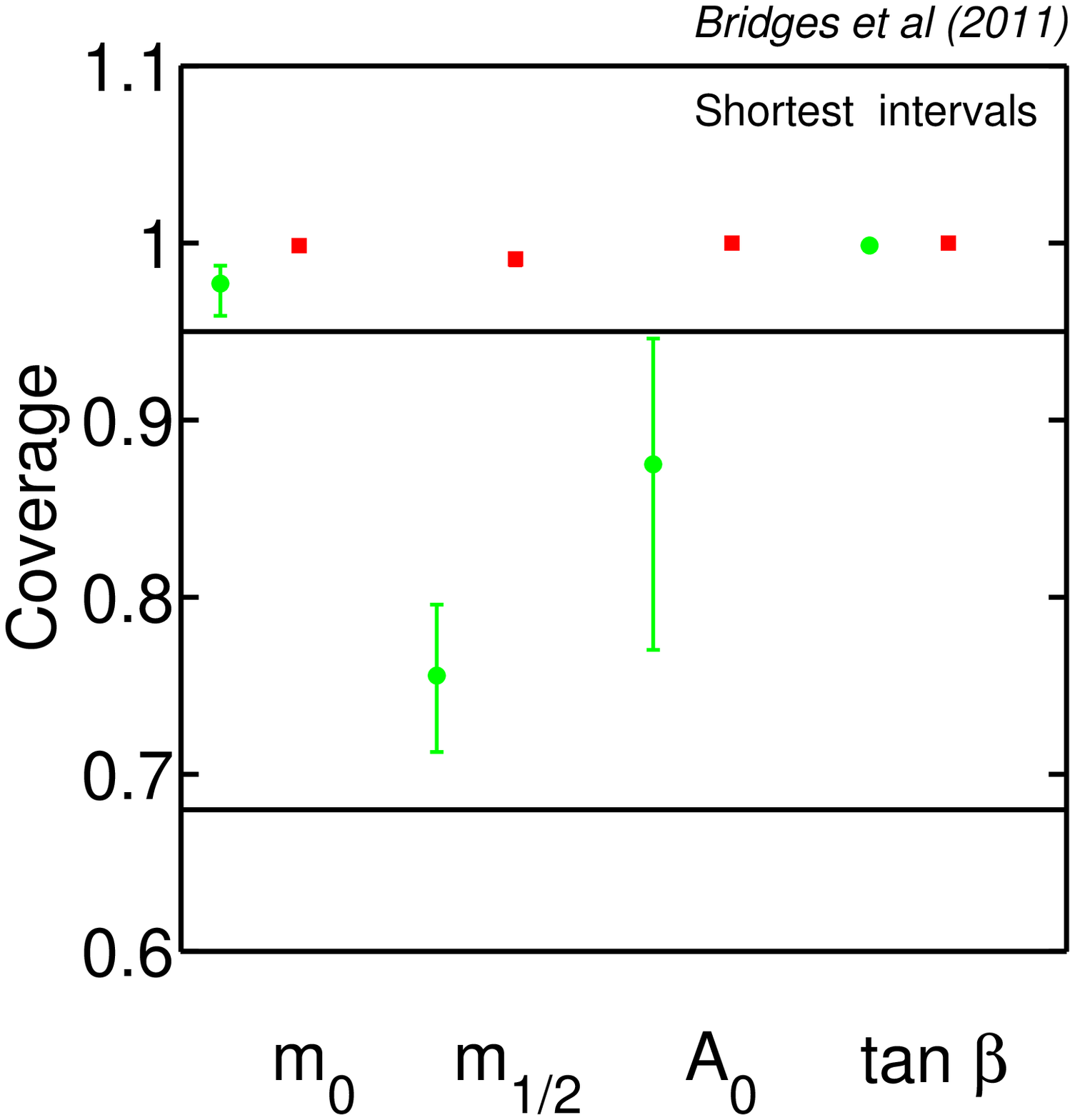} 
\includegraphics[width=0.32\linewidth]{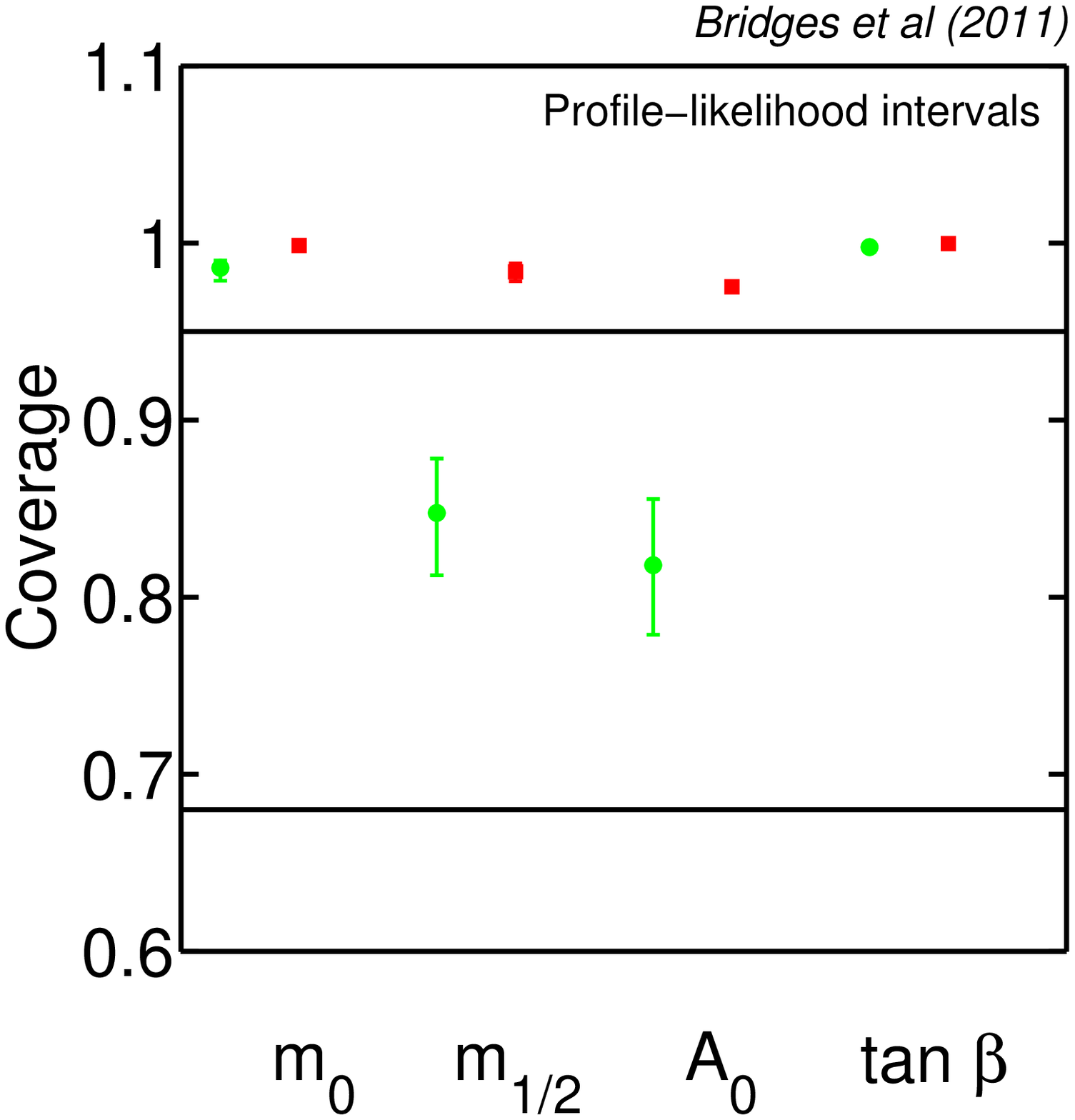}
\end{center}
\caption{\label{fig:mcmc_coverage} Coverage for various types of intervals for the CMSSM parameters, from $10^4$ realizations, employing MCMC for the reconstruction (each pseudo-experiment is reconstructed with $10^6$ samples).  Green/circles (red/squares) is for the 68\% (95\%) error. From~\cite{Bridges:2010de}.}
\end{figure}

The results are shown in Fig.~\ref{fig:mcmc_coverage}, where it can be seen that the methods have substantial over-coverage for 1-d intervals, which means that the resulting inferences are conservative. While it is difficult to unambiguously attribute the over-coverage to a specific cause, the most likely cause is the effect of boundary conditions imposed by the CMSSM. 
When ${\bf \Theta}$ is composed of parameters of interest,
$\theta$, and nuisance parameters, $\mbox{$\psi$}$, the profile likelihood ratio is defined as
\begin{equation}
\lambda(\theta) \equiv \frac{\mathcal{L}(\theta, \hat{\hat{\mbox{$\psi$}}})}{\mathcal{L}(\hat{\theta}, \hat{\mbox{$\psi$}})}.
\label{eq:profile_like}
\end{equation}
where $\hat{\hat{\mbox{$\psi$}}}$ is the conditional maximum likelihood estimate
(MLE) of $\mbox{$\psi$}$ with $\theta$ fixed and $\hat{\theta}, \hat{\mbox{$\psi$}}$ are
the unconditional MLEs. 
When the fit is performed directly in the space of the weak-scale masses (i.e., without invoking a specific SUSY model and hence bypassing the mapping ${\bf \Theta} \to {\bf m}$), there are no boundary effects, and the distribution of $-2\ln \lambda(\bf{m})$ (when $\bf{m}$ is true) is distributed as a chi-square with a number of degrees of freedom given by the dimensionality of $\bf{m}$.  Since the likelihood is invariant under reparametrizations, we expect $-2\ln \lambda(\theta)$ to also be distributed as a chi-square.    If the boundary is such that $\bf{m}(\hat{\theta},\hat{\psi}) \ne \hat{\bf{m}}$ or $\bf{m}(\theta,\hat{\hat{\psi}}) \ne \hat{\hat{\bf{m}}}$, then the resulting interval will modified.  More importantly, one expects the denominator ${\mathcal L}(\hat\theta, \hat{\psi}) < {\mathcal L}(\hat{\bf{m}} )$ since $\bf{m}$ is unconstrained, which will lead to $-2\ln \lambda(\theta) < -2\ln \lambda(\bf{m})$.  In turn, this means more parameter points being included in any given contour, which leads to over-coverage. 

The impact of the boundary on the distribution of the profile likelihood ratio is not insurmountable.  It is not fundamentally different than several common examples in high-energy physics where an unconstrained MLE would lie outside of the physical parameter space.  Examples include downward fluctuations in event-counting experiments when the signal rate is bounded to be non-negative.  Another common example is the measurement of sines and cosines of mixing angles that are physically bounded between $[-1,1]$, though an unphysical MLE may lie outside this region.  The size of this effect is related to the probability that the MLE is pushed to a physical boundary.  If this probability can be estimated, it is possible to estimate a corrected threshold on $-2\ln\lambda$.  For a precise threshold with guaranteed coverage, one must resort to a fully frequentist Neyman Construction. A similar coverage study (but without the computational advantage provided by ANNs) has been carried out for a few CMSSM benchmark points for simulated data from future direct detection experiments~\cite{YasharCoverage}. Their findings indicate substantial under-coverage for the resulting intervals, especially for certain choices of Bayesian priors. Both works clearly show the timeliness and importance of evaluating the coverage properties of the reconstructed intervals for future data sets.

\section{Challenges of profile likelihood evaluation}

For highly non-Gaussian problems like supersymmetric parameter determination, inference can depend strongly on whether one
chooses to work with the posterior distribution (Bayesian) or profile likelihood
(frequentist)~\cite{Allanach:2007qk,Trotta:2008bp,2010JCAP...01..031S}. There is a growing consensus that both the posterior
and the profile likelihood ought to be explored in order to obtain a fuller picture of the statistical constraints from
present-day and future data. This begs the question of the algorithmic solutions available to
reliably explore both the posterior and the profile likelihood in the context of SUSY phenomenology.

The profile likelihood ratio defined in Eq.~\eqref{eq:profile_like} is an attractive choice as a test statistics, for under certain regularity conditions, Wilks~\cite{Wilks}
showed that the distribution of $-2\ln\lambda(\theta)$ converges to a
chi-square distribution with a number of degrees of freedom given by
the dimensionality of $\theta$. 
 Clearly, for any given value of $\theta$, evaluation of the profile likelihood requires solving a maximisation problem in many dimensions to determine the conditional MLE $\hat{\hat{\mbox{$\psi$}}}$. While posterior samples obtained with  {\sc MultiNest} have been used to estimate the profile likelihood, the accuracy of such an estimate has been questioned~\cite{Akrami:2009hp}. As mentioned above, evaluating profile likelihoods is much more challenging than evaluating posterior
distributions. Therefore, one should not expect that a vanilla setup for {\sc MultiNest} (which is adequate for an accurate exploration of the posterior distribution) will automatically be optimal for profile likelihoods evaluation. In Ref.~\cite{Feroz:2010} the question of the accuracy of profile likelihood evaluation from {\sc MultiNest}  was investigated in detail. We report below the main results.

The two most important parameters  that control the parameter space exploration in {\sc MultiNest} are the number of live points $n_{\rm live}$ -- which determines the resolution at which the parameter space is explored --
and a tolerance parameter $\mathrm{tol}$, which defines the termination criterion based on the accuracy of the evidence. 
Generally, a larger number of live points is necessary to explore profile likelihoods accurately. Moreover, setting
$\mathrm{tol}$ to a smaller value results in {\sc MultiNest} gathering a larger number of samples in the high likelihood regions (as termination is delayed). This is usually not necessary for the posterior distributions, as the prior volume occupied by high
likelihood regions is usually very small and therefore these regions have relatively small probability mass. For
profile likelihoods, however, getting as close as possible to the true global maximum is crucial and therefore one should set
$\mathrm{tol}$ to a relatively smaller value. In Ref.~\cite{Feroz:2010} it was found that $n_{\rm live} = 20,000$ and $\mathrm{tol} = 1 \times
10^{-4}$ produce a sufficiently accurate exploration of the profile likelihood in toy models that reproduce the most important features of the CMSSM parameter space. 

\begin{figure}
\begin{center}
\includegraphics[width=1\columnwidth]{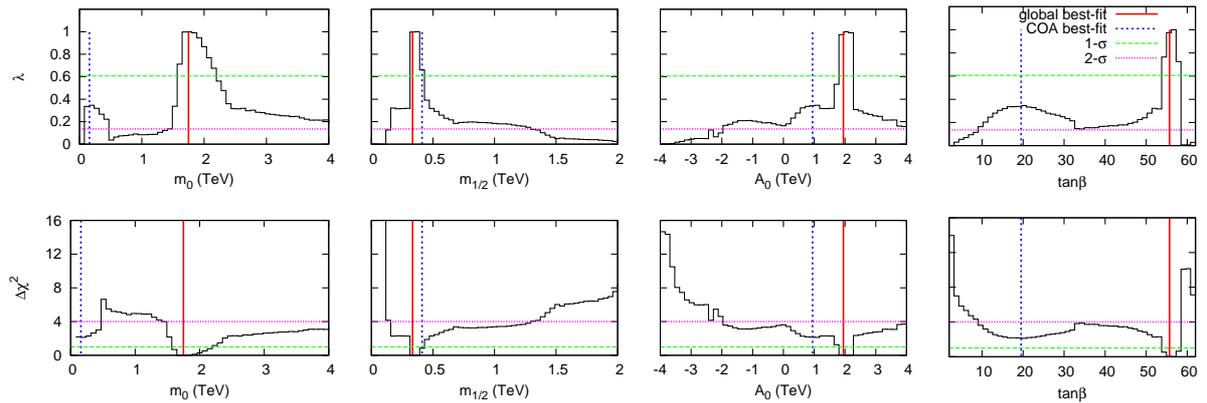}
\caption{1-D profile likelihoods from present-day data for the CMSSM parameters normalized to the global best-fit point. The red solid and blue dotted
  vertical lines represent the global best-fit point ($\chi^2 = 9.26$, located in the focus point region) and the best-fit point
  found in the stau co-annihilation region ($\chi^2 = 11.38$) respectively. The upper and lower panel show the profile
  likelihood and $\Delta\chi^2$ values, respectively. Green (magenta) horizontal lines represent the $1\sigma$ ($2\sigma$)
  approximate confidence intervals. {\sc MultiNest} was run with 20,000 live points and $\mathrm{tol}=1 \times
  10^{-4}$ (a configuration deemed appropriate for profile likelihood estimation), requiring approximately 11 million
likelihood evaluations. From \cite{Feroz:2010}. }
\label{fig:cmssm_profile_1D}
\end{center}
\end{figure}

In principle, the profile likelihood does not depend on the choice of priors. However, in order to explore the parameter space
using any Monte Carlo technique, a set of priors needs to be defined. Different choices of priors will generally lead to different regions of
the parameter space to be explored in greater or lesser detail, according to their posterior density. As a consequence, the resulting profile likelihoods might be slightly different, purely on numerical grounds.
We can obtain more robust profile likelihoods by simply merging samples obtained from scans with different choices of Bayesian priors. This does not come at a greater computational cost, given that a responsible Bayesian analysis
would estimate sensitivity to the choice of prior as well.  The results of such a scan are shown in Fig.~\ref{fig:cmssm_profile_1D}, which was obtained by tuning {\sc MultiNest} with the above configuration, appropriate for an accurate profile likelihood exploration, and by merging the posterior samples from two different choices of priors (see~\cite{Feroz:2010} for details). This high-resolution profile likelihood scan using {\sc MultiNest} compares favourably with the results obtained by adopting a dedicated Genetic Algorithm technique~\cite{Akrami:2009hp}, although at a slightly higher computational cost (a factor of $\sim 4$). In general, an accurate profile likelihood evaluation was about an order of magnitude more computationally expensive than mapping out the Bayesian posterior. 

\section{Conclusions}

As the LHC impinges on the most anticipated regions of SUSY parameter space, 
the need for statistical techniques that will be able to cope with the complexity of SUSY phenomenology
is greater than ever. An intense effort is underway to test the accuracy of parameter inference methods, both in the Frequentist and the Bayesian framework. Coverage studies such as the one presented here require highly-accelerated inference techniques, and neural networks have been demonstrated to provide a speed-up factor of up to $30,000$ with respect to conventional methods. A crucial improvement required for future coverage investigations is the ability to generate pseudo-experiments from an accurate description of the likelihood.  Both the representation of the likelihood function and the ability to generate pseudo-experiments are now possible with the workspace technology in RooFit/RooStats~\cite{RooStats}. We encourage future experiments to publish their likelihoods using this technology. Finally, an accurate evaluation of the profile likelihood remains a numerically challenging task, much more so than the mapping out of the Bayesian posterior. Particular care needs to be taken in tuning appropriately Bayesian algorithms targeted to the exploration of posterior mass (rather than likelihood maximisation). We have demonstrated that the  {\sc MultiNest} algorithm can be succesfully employed for approximating the profile likelihood functions, even
though it was primarily designed for Bayesian analyses.  
  In particular, it is important to use a
termination criterion that allows {\sc MultiNest} to explore
high-likelihood regions to sufficient resolution.

{\em Acknowledgements:} We would like to thank the organizers of PHYSTAT11 for a very interesting workshop. We are grateful to Yashar Akrami, Jan Conrad, Joakim Edsj\"o, Louis Lyons and Pat Scott for many useful discussions.

\end{document}